\documentclass[%
prx
,twocolumn%
,amsmath,amssymb,aps,nobibnotes
,notitlepage
,nofootinbib
,preprintnumbers
]{revtex4-1}

\usepackage[colorlinks,linkcolor=blue,citecolor=blue]{hyperref}
\usepackage{graphicx}
\usepackage{bm}
\usepackage{braket}
\usepackage{color}
\usepackage{tabularx}
\usepackage{mathtools}

\begin{document}
\title{Berry Phase and Topology in Ultrastrongly Coupled Quantum Light-Matter Systems}
\author{Kanta Masuki}
\email{masuki@g.ecc.u-tokyo.ac.jp}
\affiliation{Department of Physics, University of Tokyo, 7-3-1 Hongo, Bunkyo-ku, Tokyo 113-0033, Japan}

\author{Yuto Ashida}
\email{ashida@phys.s.u-tokyo.ac.jp}
\affiliation{Department of Physics, University of Tokyo, 7-3-1 Hongo, Bunkyo-ku, Tokyo 113-0033, Japan}
\affiliation{Institute for Physics of Intelligence, University of Tokyo, 7-3-1 Hongo, Bunkyo-ku, Tokyo 113-0033, Japan}

\begin{abstract}
  Strong coupling between matter and quantized electromagnetic fields in a cavity has emerged as a possible route toward controlling the phase of matter in the absence of an external drive. We develop a faithful and efficient theoretical framework to analyze quantum geometry and topology in materials ultrastrongly coupled to cavity electromagnetic fields in two dimensions. The formalism allows us to accurately evaluate geometrical and topological quantities, such as Berry phase and Chern number, in ultrastrong and deep strong coupling regimes. We apply our general framework to analyze a concrete model of massive Dirac fermions coupled to a circularly polarized cavity mode. Surprisingly, in addition to an ordinary transition to the topological phase, our analysis reveals a qualitatively new feature in deep strong coupling regimes, namely, the emergence of reentrant transition to the topologically trivial phase.  We also present its intuitive understanding by showing the unitary mapping between the low-energy effective theory of strongly coupled light-matter systems and the Haldane honeycomb model.
\end{abstract}

\maketitle

\section{Introduction\label{sec:intro}}
Berry phase plays a central role in modern condensed matter physics  \cite{berry_quantal_1983,provost_riemannian_1980,vanderbilt_berry_2018}. It lies at the heart of a wide variety of intriguing phenomena, including electric polarization of crystalline insulators \cite{resta_theory_1992,*resta_macroscopic_1994,vanderbilt_berry_2018}, the anomalous Hall effect \cite{klitzing_new_1980,thouless_quantized_1982,kane_quantum_2005,sheng_quantum_2006,nagaosa_anomalous_2010}, and electromagnetic responses \cite{tarruell_creating_2012,rhim_quantum_2020,hwang_geometric_2021,topp_light-matter_2021,ahn_riemannian_2022}. In particular, the last few decades have witnessed significant advances in our understanding of the intimate connection between Berry phase and topological phases of matter \cite{hasan_colloquium_2010,qi_topological_2011,thouless_quantization_1983,haldane_model_1988,fu_time_2006,moore_topological_2007,schnyder_classification_2008,kitaev_periodic_2009,teo_topological_2010}. Besides,  ongoing experimental developments in ultracold atoms \cite{schafer_tools_2020,lohse_thouless_2016,goldman_topological_2016,aidelsburger_experimental_2011,jotzu_experimental_2014,dudarev_spin-orbit_2004,celi_synthetic_2014,song_observation_2018,asteria_measuring_2019} and photonics  \cite{hafezi_robust_2011,kitagawa_observation_2012,lu_topological_2014,wu_scheme_2015,maczewsky_observation_2017,mukherjee_experimental_2017,ozawa_topological_2019,gianfrate_measurement_2020} have allowed one to study these rich phenomena in a highly controllable way.

On another front, recent developments have made it possible to realize strong interactions between matter and quantized electromagnetic fields inside a cavity \cite{anappara_signatures_2009,scalari_ultrastrong_2012,maissen_ultrastrong_2014,gambino_exploring_2014,chikkaraddy_single-molecule_2016,bayer_terahertz_2017,halbhuber_non-adiabatic_2020,genco_bright_2018,YF172,flick_atoms_2017,forn-diaz_ultrastrong_2019,frisk_kockum_ultrastrong_2019}. In particular, the emergent field of cavity quantum electrodynamics (QED) materials has attracted significant attention as a possible platform for controlling the phase of matter in the absence of an external drive \cite{garcia-vidal_manipulating_2021,hubener_engineering_2021,owens_chiral_2022,schlawin_cavity_2022,BJ22}. So far, the effects of cavity confinement on chemical reactions \cite{ruggenthaler_quantum-electrodynamical_2018,ribeiro_polariton_2018,flick_strong_2018,flick_lightmatter_2019,galego_cavity-induced_2015,herrera_cavity-controlled_2016,chervy_high-efficiency_2016,ebbesen_hybrid_2016,feist_polaritonic_2018,thomas_ground-state_2016,stranius_selective_2018,martinez-martinez_polariton-assisted_2018,martinez-martinez_triplet_2019,polak_manipulating_2020,xiang_intermolecular_2020} and such diverse phenomena as superconductivity \cite{scalari_superconducting_2014,schlawin_cavity-mediated_2019,curtis_cavity_2019,curtis_cavity_2022,gao_higgs_2021,sentef_cavity_2018}, ferroelectricity \cite{ashida_quantum_2020,PP20,latini_ferroelectric_2021,LK22}, and the quantum Hall effect \cite{smolka_cavity_2014,ciuti_cavity-mediated_2021,rokaj_quantum_2019,rokaj_polaritonic_2022,appugliese_breakdown_2022} have been experimentally and theoretically investigated \cite{zhang_collective_2016,paravicini-bagliani_magneto-transport_2019,keller_landau_2020,MNS20,orgiu_conductivity_2015,kiffner_manipulating_2019,*kiffner_mott_2019,li_effective_2022,tokatly_vacuum_2021,wang_cavity_2019,kibis_band_2011,ashida_cavity_2021,ashida_nonperturbative_2022,li_electromagnetic_2020,chiocchetta_cavity-induced_2021,mann_tunable_2020,mann_manipulating_2018,juraschek_cavity_2021,hagenmuller_cavity-enhanced_2017,hagenmuller_cavity-assisted_2018,bartolo_vacuum-dressed_2018,amelio_optical_2021,andolina_cavity_2019,downing_topological_2019,downing_quantum_2022,karzig_topological_2015,knorr_intersubband_2022,ravets_polaron_2018,klembt_exciton-polariton_2018,konishi_universal_2021,chervy_accelerating_2020,le_de_cavity_2022,mazza_superradiant_2019,roman-roche_effective_2022,thomas_large_2021,NC2010,GS2017,MK2013,BP2016,L2014}.

To further explore the potential of cavity QED engineering, the time is ripe to examine how strong light-matter interactions can influence the quantum geometrical and topological properties of matter. While related problems were often addressed by resorting to simplified phenomenological descriptions, such as Peierls substitution or minimally coupled effective Hamiltonians, the validity of  these conventional approaches becomes questionable \cite{schlawin_cavity_2022}  in ultrastrong coupling (USC) and  deep strong coupling (DSC) regimes that are within reach of recent experiments \cite{anappara_signatures_2009,scalari_ultrastrong_2012,maissen_ultrastrong_2014,gambino_exploring_2014,chikkaraddy_single-molecule_2016,bayer_terahertz_2017,halbhuber_non-adiabatic_2020,genco_bright_2018,YF172,flick_atoms_2017,forn-diaz_ultrastrong_2019,frisk_kockum_ultrastrong_2019,keller_landau_2020,MNS20,halbhuber_non-adiabatic_2020,appugliese_breakdown_2022,orgiu_conductivity_2015}. Thus, the lack of faithful and efficient theoretical approach to study  quantum geometry and topology in ultrastrongly coupled systems remains a major current issue.

\begin{figure}[b]
  \includegraphics[width = 8.0cm]{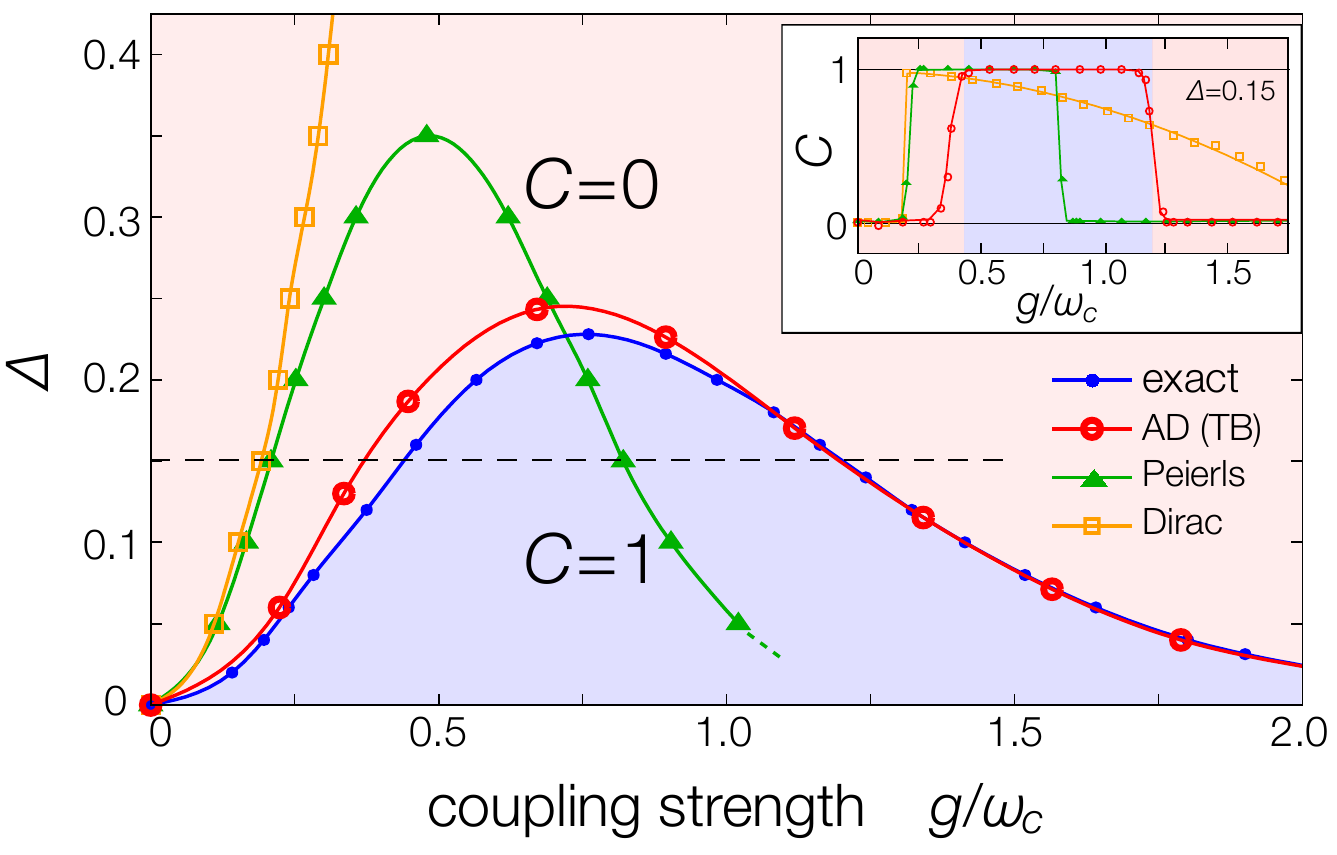}
  \caption{\label{figure1}Phase diagram of massive Dirac fermions coupled to circularly polarized cavity photons (cf.~Eq.~\eqref{ham_Coulomb} with the honeycomb lattice potential~\eqref{potential_massive_honeycomb}). The phase boundary is determined by the Chern number \(C\) of the lowest electron-polariton band. At a small on-site energy difference \(\Delta\), the light-matter coupling \(g\) first gives rise to a transition from the trivial phase (\(C\!=\!0\)) to the topological phase (\(C\!=\!1\)), and then exhibits the reentrant behavior in DSC regimes (\(g/\omega_c\!>\!1\)).
    We compare the phase boundaries obtained from different methods, including  the exact analysis of Eq.~\eqref{ham_AD} (blue), the tight-binding model~\eqref{ham_TB} constructed in the AD frame (red), the tight-binding model with  Peierls substitution (green), and the minimally coupled Dirac Hamiltonian (orange). The inset shows the corresponding  Chern numbers at \(\Delta\!=\!0.15\). We take the unit of \(\hbar = q = a = 1\) throughout this paper. Parameters are \(m\!=\!5, \omega_c\!=\!4, V_0\!=\!1.5\).}
\end{figure}

The aim of this paper is to develop a theoretical framework to accurately analyze geometrical and topological properties of matter coupled to cavity electromagnetic fields. To this end, we start from a general single-electron Hamiltonian with a single cavity mode (see Eq.~\eqref{ham_Coulomb} below). After employing the asymptotically decoupling (AD) unitary transformation, we derive an expression of Berry curvature (Eq.~\eqref{Berry_AD})  that allows for its efficient evaluation at arbitrary coupling strengths on the basis of a simple tight-binding description. Our formalism thus provides a faithful and efficient way to evaluate Berry phase and related topological numbers in generic two-dimensional cavity QED materials.

To demonstrate the advantage of the present formalism, we apply it to a concrete model of massive Dirac fermions on honeycomb lattice potential, which can exhibit cavity-induced topological phase transition (TPT). Figure~\ref{figure1} shows the phase diagram determined by the Chern number \(C\) of the lowest electron-polariton band. In the absence of the light-matter interaction (\(g\!=\!0\)), a nonzero on-site energy difference \(\Delta\) breaks the inversion symmetry, leading to the gap opening at the Dirac points and the trivial phase with \(C\!=\!0\) (see also Fig.~\ref{figure2} below). As the coupling strength \(g\) is increased, the transition to the topological phase with \(C\!=\!1\) can occur due to the time-reversal-symmetry (TRS) breaking inherent to the circularly polarized cavity mode. Interestingly, when the coupling \(g\) is further increased, the system exhibits the topological-to-trivial reentrant behavior in DSC regimes \(g/\omega_c\!\gtrsim\!1\). We show that this nonmonotonic behavior can be understood from the connection between the low-energy effective theory of the present light-matter Hamiltonian and the Haldane honeycomb model \cite{haldane_model_1988}.

The rest of the paper is organized as follows. In Sec.~\ref{sec:Model}, we start from the cavity QED Hamiltonian in the Coulomb gauge and construct the effective model in the AD frame. Then, in Sec.~\ref{sec:Berry_curv_in_AD}, we provide the exact expression of Berry curvature in the AD frame~\eqref{Berry_AD} and its useful approximation. In Sec.~\ref{sec:Honeycomb}, we apply our formalism to a concrete model of Dirac electrons coupled to a circularly polarized cavity mode, and discuss the topological phase transitions in this system. We also perform thorough comparative analyses of geometrical and topological quantities, demonstrating the reliability of our approach compared to the commonly used methods in the literature (see Figs.~\ref{figure3} and \ref{figure4}). Section~\ref{sec:Discussion} summarizes the results and discusses the qualitative understandings of the topological phase transitions. Some technical details are discussed in Appendix.

\section{Model Description\label{sec:Model}}
To illustrate our formalism in a simple yet experimentally relevant setup where Berry curvature can be nontrivial, we consider an electron interacting with a  cavity mode in two dimensions, while the following discussions can be generalized to systems with multiple cavity modes in arbitrary dimensions. For the sake of  simplicity, we neglect the electron spin degrees of freedom. We also assume that the photon loss is sufficiently small and consider the Hamiltonian under the long-wavelength approximation, which is given in the Coulomb gauge as~\cite{cohen-tannoudji_photons_1989,loudon_quantum_2000}
\begin{align}
  \hat H_C    & = \frac{(\hat{\bm p}-q\hat{\bm A})^2}{2m} + V(\bm r) + \hbar\omega_c\left(\hat a^\dagger\hat a+\frac{1}{2}\right),\label{ham_Coulomb} \\
  \hat{\bm A} & = A_0\left(\bm f\hat a + \bm f^*\hat a^\dagger\right)\label{vector_potential},\ \bm f\in \mathbb{C}^2.
\end{align}
Here, \(m\) and \(q\) are the mass and the charge of an electron, respectively,  \(V(\bm r)\!= \!\sum_{\bm G} V_{\bm G} e^{i\bm G\bm r}\) is an arbitrary two-dimensional periodic potential, and \(\hat a^{(\dagger)}\) is the annihilation (creation) operator of cavity photons with frequency \(\omega_c\). An electron couples to cavity photons through the vector potential \(\hat{\bm A}\) in Eq.~\eqref{vector_potential}, where \(A_0\) is the mode amplitude, and \(\bm f\) is an arbitrary polarization vector that satisfies \(|\bm f| = 1\). The coupling strength of this light-matter interaction is characterized by
\begin{align}
  g = qA_0\sqrt{\frac{\omega_c}{m\hbar}}.
\end{align}
It is the common convention that the system is said to be in ultrastrong coupling regimes for \(0.1\leq g/\omega_c \leq 1\) and deep strong coupling regimes for \(g/\omega_c \geq 1\).
We note that the cavity QED Hamiltonian~\eqref{ham_Coulomb} possesses TRS if and only if there exists \(\theta\in \mathbb{R}\) such that the polarization vector \(\bm f\) satisfies \(\bm f\!=\! e^{i\theta}\bm f^*\). To examine the physically interesting case with broken TRS, we set \(\bm f\) as a circularly polarized vector \(\bm e = (1,-i)^T/\sqrt{2}\) in the following. See Appendices~\ref{App:AD_transformation} and~\ref{App:AD_multi} for the generalization to systems with arbitrary polarization or multiple electromagnetic modes. We expect that TRS-breaking cavity modes can be realized in chiral~\cite{owens_chiral_2022,hubener_engineering_2021} or  gyrotropic cavities~\cite{feigh_eigenmode_2004,tokatly_vacuum_2021}.

The Hamiltonian~\eqref{ham_Coulomb} possesses the discrete translational symmetry of \(V(\bm r)\), which allows us to label its eigenstate as Bloch state \(|\psi_{n\bm k}^C\rangle\) with band index \(n\) and Bloch wavevector \(\bm k\). In strong coupling regimes, \(|\psi_{n\bm k}^C\rangle\) is in general a highly entangled electron-photon state.  A wide variety of experimental observables, including electromagnetic responses and transport coefficients, can be related to Berry curvature of these Bloch states:
\begin{align}
  B^C_{n}(\bm k) = i\bigl( \langle \partial_{k_x} u^C_{n\bm k}|\partial_{k_y} u^C_{n\bm k}\rangle - \langle \partial_{k_y} u^C_{n\bm k}|\partial_{k_x} u^C_{n\bm k}\rangle\bigr),\label{Berry_Coulomb}
\end{align}
where \(|u_{n\bm k}^C\rangle\) is the periodic part of Bloch state defined by \(|u_{n\bm k}^C\rangle = e^{-i\bm k\bm r}|\psi_{n\bm k}^C\rangle\). We again emphasize that \(|u_{n\bm k}^C\rangle\) is not a mere electron state but an entangled wavefunction on the Hilbert space spanned by the product of electron and photon states.
In USC and DSC regimes, the calculation of Berry curvature~\eqref{Berry_Coulomb} by the exact diagonalization of Eq.~\eqref{ham_Coulomb} becomes eventually intractable due to the need of including increasingly many electron-photon states in this frame.
Instead, previous studies often resort to phenomenological descriptions  such as  Peierls substitution or minimally coupled effective Hamiltonians (see, e.g., Refs.~\cite{kiffner_manipulating_2019,wang_cavity_2019,kibis_band_2011}). However, as detailed below, these simplified methods become invalid at strong light-matter couplings.
To accurately evaluate Berry curvature~\eqref{Berry_Coulomb} and related topological numbers in an efficient way, we thus need to compress the dimension of the relevant Hilbert space while keeping nontrivial effects of light-matter interactions without making ambiguous simplifications.

To achieve this, we perform the AD transformation~\cite{ashida_cavity_2021}
\begin{align}
  \hat U        & = \exp\left(-i\xi\frac{\hat{\bm p}}{\hbar}\cdot\hat{\bm \pi}\right),            \\
  \xi           & = \sqrt{\frac{\hbar\omega_c}{m}} \cdot \frac{g}{\omega_c^2 + g^2},\label{xi_AD} \\
  \hat{\bm \pi} & = -i\bm e\hat a + i\bm e^*\hat a^\dagger.
\end{align}
The transformation \(\hat U\) acts on the creation operator~\(\hat a^\dagger\) and the position operator \(\bm r\) via~\(\hat U^\dagger \hat a^\dagger \hat U\!=\!\hat a^\dagger\!+\!\xi\hat{\bm p}\!\cdot\!\bm e/\hbar\) and \(\hat U^\dagger\bm r\hat U \!=\! \bm r\!+\!\xi\hat{\bm \pi}\!+\!\xi^2\hat{\bm p}\!\times\!\bm e_z/(2\hbar)\), respectively. We then obtain the transformed Hamiltonian \(\hat H^U \equiv \hat U^\dagger\hat H\hat U\) as
\begin{align}
  \hat H^U\!  & =\! \frac{\hat{\bm p}^2}{2m_{\rm eff}}+V\!\left(\bm r+\xi\hat{\bm \pi}\!+\!\frac{\xi^2}{2\hbar}\hat{\bm p}\times\bm e_z\right)\!+\!\hbar\Omega\left(\hat a^\dagger\hat a+\frac{1}{2}\right),\label{ham_AD} \\
  m_{\rm eff} & = m\left( 1 + \frac{g^2}{\omega_c^2} \right),\ \Omega = \omega_c\left( 1 + \frac{g^2}{\omega_c^2} \right).
\end{align}
Here, \(m_{\rm eff}\) is the effective mass dressed by cavity photons, and \(\Omega\) is the renormalized cavity frequency due to the \(\hat{\bm A}^2\)-term in the Coulomb gauge~\eqref{ham_Coulomb}. In this frame, the light-matter interaction is expressed as the shift of the electron coordinate \(\bm r\), and its effective strength is characterized by the parameter \(\xi\) defined in Eq.~\eqref{xi_AD}. Because \(\xi\) behaves as \(\xi \propto g\) at small \(g\) and \(\xi \propto g^{-1}\) at large \(g\), light and matter degrees of freedom are asymptotically decoupled in the weak and strong coupling limits. One can thus well approximate the low-energy eigenstates simply by the product state of an electron wavefunction and the photon vacuum \(|0\rangle\). This feature allows us to construct a useful low-energy effective description in the following way. Firstly, we construct Wannier orbitals \(|w_i\rangle\) from the matter part of $\hat{H}^U$:
\begin{align}
  \hat H_{\rm mat}^U\!=\!\langle 0|\hat H^U|0\rangle\!=\!\frac{\hat{\bm p}^2}{2m_{\rm eff}}\!+\!\sum_{\bm G} V_{\bm G}e^{-\frac{\xi^2G^2}{4}} e^{i\bm G\cdot\left(\bm r+\frac{\xi^2}{2\hbar}\hat{\bm p}\times\bm e_z\right)},\label{ham_mat}
\end{align}
where \(|0\rangle\) is the photon vacuum, and \(V_{\bm G}\) is the Fourier component of the periodic potential \(V(\bm r)\) defined via \(V(\bm r) \!=\! \sum_{\bm G} V_{\bm G} e^{-i\bm G\cdot\bm r}\). Secondly, we project the continuum Hamiltonian \(\hat H^U\) onto the manifold spanned by these Wanneir orbitals, obtaining the following tight-binding Hamiltonian:
\begin{align}
  \hat H^U_{\rm TB} = \sum_{\langle ij \rangle} (t_{ij} + \hat \mu_{ij})|w_i\rangle\langle w_j|,\label{ham_TB}
\end{align}
where \(t_{ij} = \langle w_i|\hat H^U_{\rm mat}|w_j\rangle\) is an effective hopping amplitude, and \(\hat \mu_{ij} = \langle w_i|(\hat H^U-\hat H^U_{\rm mat})|w_j\rangle\) represents electromagnetic fluctuations.

In the newly obtained tight-binding Hamiltonian~\eqref{ham_TB}, the underlying Wannier orbitals \(|w_i\rangle\) already incorporate light-matter interactions in a nonperturbative manner through the renormalized mass and potential in Eq.~\eqref{ham_mat}. The hopping amplitudes $t_{ij}$ play the dominant role while the residual terms \(\hat \mu_{ij}\) are typically negligible and do not qualitatively alter the results, as shown in Appendix~\ref{App:EM_fluctuations}.
We note that our construction of the tight-binding model makes a sharp contrast with usual phenomenological methods; for instance,  in Peierls substitution, one starts from a tight-binding Hamiltonian \(\sum_{\langle ij\rangle} t_{ij}^0|w_i^0\rangle\langle w_j^0|\) with the bare Wannier orbitals $|w_i^0\rangle$ in the absence of light-matter interactions and then replaces the hopping amplitudes \(t_{ij}^0\) by \(t_{ij}^0 \exp[i(q/\hbar)\int_{\bm r_j}^{\bm r_i}d\bm r\cdot\! \hat{\bm A}]\) with \(\bm r_{i}\!=\!\langle w_{i}^0|\bm r|w_{i}^0\rangle\).

\section{Berry curvature in the transformed frame\label{sec:Berry_curv_in_AD}}
We recall that physical observables can be related to Berry curvature~\eqref{Berry_Coulomb}, which is originally defined in the Coulomb gauge. To derive the exact expression of Berry curvature~\eqref{Berry_Coulomb} after the unitary transformation, we note that the discrete translational symmetry remains intact in the transformed Hamiltonian~\eqref{ham_AD}, i.e.,  eigenstates in this frame are also Bloch states \(|\psi_{n\bm k}^U\rangle = e^{i\bm k\bm r}|u_{n\bm k}^U\rangle\) and can be related to their Coulomb-gauge counterparts by \(|u_{n\bm k}^U\rangle = \exp[\frac{i\xi}{\hbar}(\hat{\bm p}+\hbar\bm k)\!\cdot\!\hat{\bm \pi}]|u_{n\bm k}^C\rangle\).
We can then rewrite Eq.~\eqref{Berry_Coulomb} in terms of the transformed eigenstates as
\begin{align}
  B^C_{n}(\bm k) = & i\left(\langle\partial_{k_x} u_{n\bm k}^U|\partial_{k_y} u_{n\bm k}^U\rangle - \langle\partial_{k_y} u_{n\bm k}^U|\partial_{k_x} u_{n\bm k}^U\rangle\right)\nonumber                                                                            \\
  +                & \xi \left[\bm \nabla_{\bm k}\!\times\!\langle\psi_{n\bm k}^U|\hat{\bm \pi}|\psi_{n\bm k}^U\rangle\right]_{k_z} \!\!-\! \frac{\xi^2}{2\hbar}\bm\nabla_{\bm k}\!\cdot\!\langle\psi_{n\bm k}^U|\hat{\bm p}|\psi_{n\bm k}^U\rangle,\label{Berry_AD}
\end{align}
where the latter two terms originate from the \(\bm k\)-dependence of the unitary transformation between \(|u_{n\bm k}^C\rangle\) and \(|u_{n\bm k}^U\rangle\). See Appendix~\ref{App:quantum_geometry} for derivations.

The dominant contribution in Eq.~\eqref{Berry_AD} comes from the terms in the first line, while the other terms can be neglected owing to the asymptote $\xi\!\propto\!g^{-1}$ at large \(g\). Together with the fact that the  electromagnetic fluctuations \(\hat\mu_{ij}\) in Eq.~\eqref{ham_TB} can also be neglected, the analysis of Berry phase in the transformed frame can significantly be simplified as follows:
\begin{align}
  B^C_{n}(\bm k) \simeq i\left(\langle\partial_{k_x} \tilde u_{n\bm k}^{U}|\partial_{k_y} \tilde u_{n\bm k}^{U}\rangle - \langle\partial_{k_y} \tilde u_{n\bm k}^{U}|\partial_{k_x} \tilde u_{n\bm k}^{U}\rangle\right),\label{Berry_AD_approx}
\end{align}
where \(|\tilde u_{n\bm k}^{U}\rangle\) is an electron eigenstate of \(\sum_{\langle ij\rangle} t_{ij}|w_i\rangle\langle w_j|\), which only contains the matter part of the tight-binding Hamiltonian~\eqref{ham_TB}.
One can thus accurately and efficiently evaluate Berry curvature of ultrastrongly coupled systems simply by Eq.~\eqref{Berry_AD_approx}  without photon degrees of freedom. Instead, it is also possible to include higher-order terms to systematically improve quantitative accuracy when necessary. The same strategy is applicable to calculations of other geometric quantities such as Berry connection and quantum metric as described in Appendix~\ref{App:quantum_geometry}.

\begin{figure}[t]
  \includegraphics[width = 8.5cm]{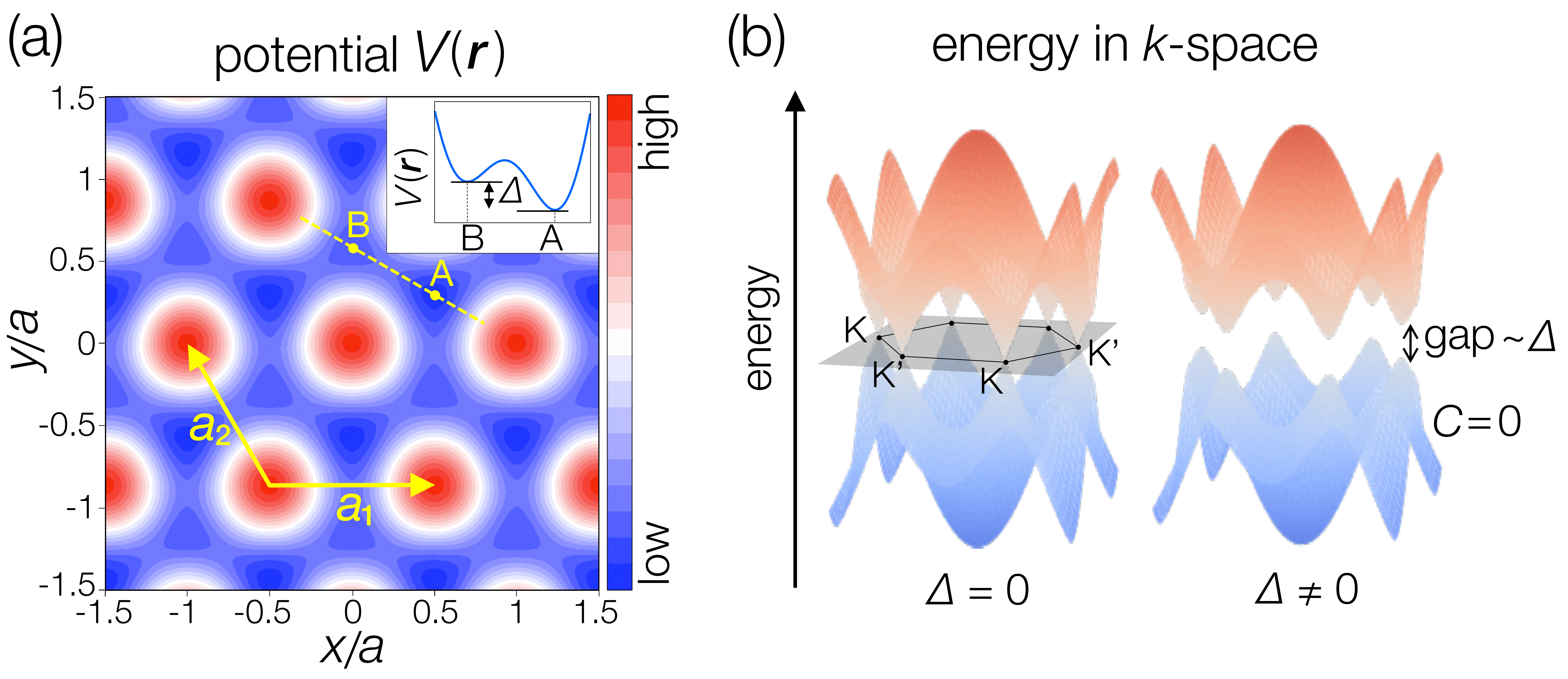}
  \caption{\label{figure2}(a) Honeycomb lattice potential \(V(\bm r)\) in Eq.~\eqref{potential_massive_honeycomb}. The inset shows the potential along the yellow dashed line, where \(\Delta\) corresponds to the energy difference between the two sublattices. (b) Two lowest energy bands in the absence of cavity fields. Nonzero \(\Delta\) opens an energy gap in the Dirac cones, where the lowest band is trivial with the Chern number \(C=0\).}
\end{figure}

\section{Honeycomb model\label{sec:Honeycomb}}
To be concrete, we consider a model of massive Dirac fermions on honeycomb lattice potential, that is, the light-matter Hamiltonian~\eqref{ham_Coulomb} with the following potential:
\begin{align}
  V(\bm r)    & = V_0\phi(\bm r) + \frac{\Delta}{9}\phi\left(\bm r- \frac{\bm a_1}{3} + \frac{\bm a_2}{3}\right),\label{potential_massive_honeycomb} \\
  \phi(\bm r) & = \sum_{\bm G= \pm \bm b_1, \pm \bm b_2, \pm(\bm b_1-\bm b_2)} \exp(i\bm G\cdot\bm r),
\end{align}
where \(\bm a_1 = (a,0)\) and \(\bm a_2 = (-a/2, \sqrt{3}a/2)\) are the primitive lattice vectors with their reciprocal counterparts \(\bm b_1 = (2\pi/a, 2\pi/\sqrt{3}a)\) and \(\bm b_2 = (0, 4\pi/\sqrt{3}a)\). As shown in Fig.~\ref{figure2}(a), the minima of this potential are aligned on the honeycomb lattice, and the on-site energy difference between the two sublattices is given by $\Delta$.

Figure~\ref{figure2}(b) shows the two lowest energy bands in the absence of cavity electromagnetic fields. A nonzero \(\Delta\) breaks the inversion symmetry and opens an energy gap \(\sim\!\Delta\) in the Dirac cones  \cite{manes_existence_2007}, resulting in the topologically trivial lowest energy band with the Chern number \(C \!=\! 0\).
When the light-matter coupling is included, the TRS-breaking by the circularly polarized cavity mode is expected to trigger the eventual gap closing and the transition to the topologically nontrivial phase with \(C \!=\! \pm 1\). While such a cavity-induced TPT was previously explored within a minimally coupled Dirac Hamiltonian~\cite{wang_cavity_2019,kibis_band_2011}, this simplified analysis can be justified only in perturbative regimes \(g/\omega_c\!\ll\!1\) as demonstrated below.

\begin{figure}[t]
  \centering
  \includegraphics[width = 8cm]{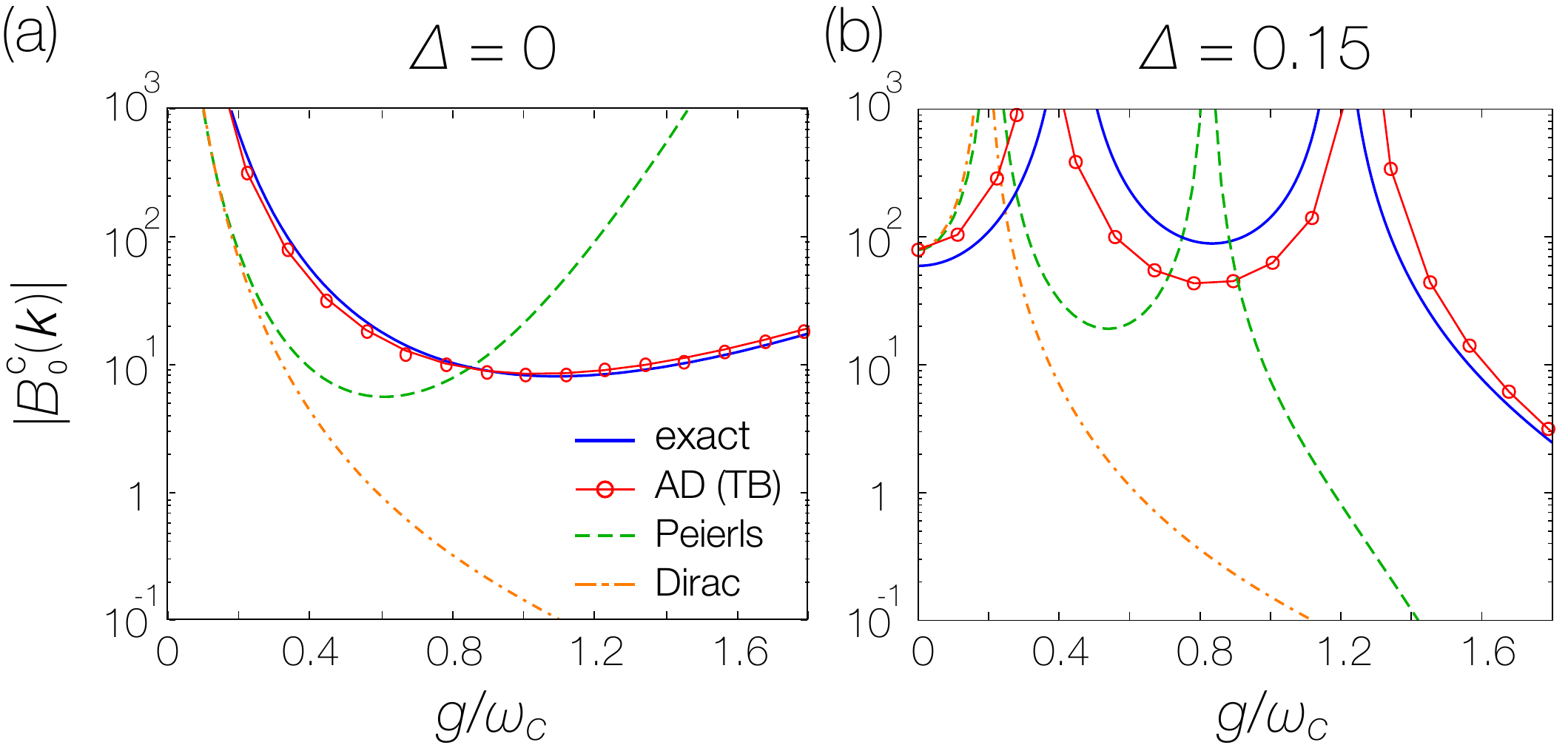}
  \caption{\label{figure3}(a), (b) Absolute value of the Berry curvature of the lowest electron-polariton band at the Dirac point at different \(\Delta\) in Eq.~\eqref{potential_massive_honeycomb}. We compare the results obtained from  the exact Hamiltonian~\eqref{ham_AD} with Eq.~\eqref{sm_Berry_exact}, which is equivalent to Eq.~\eqref{Berry_AD} (blue), the electron-only tight-binding model with Eq.~\eqref{Berry_AD_approx} (red), the tight-binding model with Peierls substitution (green), and the minimally coupled Dirac Hamiltonian~\eqref{sm_ham_min_Dirac} (orange). Parameters are \(m\!=\!5, \omega_c\!=\!4, V_0\!=\!1.5, \Delta\!=\!0\).}
\end{figure}

We now examine the low-energy properties of the present model on the basis of our formalism; hereafter we take the unit of \(\hbar\!=\!q\!=\!a\!=\!1\). To construct the effective tight-binding model~\eqref{ham_TB}, we use the maximally localized Wannier orbitals  \(|w_i\rangle\)   of Eq.~\eqref{ham_mat} \cite{marzari_maximally_1997,souza_maximally_2001,marzari_maximally_2012,pizzi_wannier90_2020}
and consider up to the next nearest-neighbor hoppings on the honeycomb lattice. We then analyze the behavior of Berry curvature, Chern number, and the corresponding energy spectrum with a varying coupling $g$, and compare the results with the ones obtained from other methods. For instance, the results of minimally coupled Dirac Hamiltonian are obtained with Eq.~\eqref{sm_ham_min_Dirac}. These results are compared to the exact results obtained by the exact diagonalization of the full continuum Hamiltonian~\eqref{ham_AD} and using the expression of Berry curvature~\eqref{sm_Berry_exact}, which is equivalent to Eq.~\eqref{Berry_AD}. We note that the exact analyses in general require high computational costs, and it is crucial to develop an effective description to efficiently evaluate expectation values of physical observables as done here. See Appendices~\ref{App:minimally_coupled} and~\ref{App:exact} for further details.

Figure~\ref{figure1} shows the obtained phase diagram whose phase boundary is determined by the Chern number \(C\) of the lowest electron-polariton band, i.e., \(C\!=\!(2\pi)^{-1}\int_{\rm BZ} d^2\bm k\ B_0^C(\bm k)\) \footnote{As for the minimally coupled Dirac Hamiltonian, we determine the phase boundary by locating \(g\) at which the band gap closes.}. At a small energy difference \(\Delta \lesssim 0.2\), TPTs from the trivial phase (\(C\!=\!0\)) to the topological phase (\(C\!=\!1\)) can occur in USC regimes. Surprisingly, the system also exhibits the reentrant transition to the trivial phase in DSC regimes \(g/\omega_c\!\gtrsim\!1\), whose interpretation will be given below. It is also worthwhile to emphasize that TPTs no longer exist at a sufficiently large $\Delta$. The minimally coupled Dirac Hamiltonian (orange curve) fails to capture these intriguing features and, more importantly, it even gives an unquantized Chern number (see the inset). Peierls substitution also fails to give quantitatively accurate results once the USC regime \(g/\omega_c\!\gtrsim\!0.1\) is reached. In contrast, our tight-binding model~\eqref{ham_TB} (red curve) remains valid over the whole parameter region at both qualitative and quantitative levels.

To make further comparisons, we plot the absolute value of the Berry curvature of the lowest electron-polariton band at the K-point, i.e., \(\bm k = \bm b_1/3 + \bm b_2/3\), with a varying coupling $g$ (Fig.~\ref{figure3}). We note that the Berry curvature at the K-point takes a positive (negative) value in the topological (topologically trivial) phase. Again, our results (red curve) obtained from the formula~\eqref{Berry_AD_approx} associated with the electron-only tight-binding model, \(\sum_{\langle ij\rangle} t_{ij}|w_i\rangle\langle w_j|\), agree well with the exact results (blue curve) at arbitrary coupling strengths, while our results slightly deviate from the exact ones in the vicinity of the band-gap closing points (Fig.~\ref{figure3}b). Meanwhile, the other conventional descriptions dramatically fail in USC and DSC regimes, where Berry curvature is either under- or overestimated by orders of magnitude. 

\begin{figure}[t]
  \centering
  \includegraphics[width = 8.5cm]{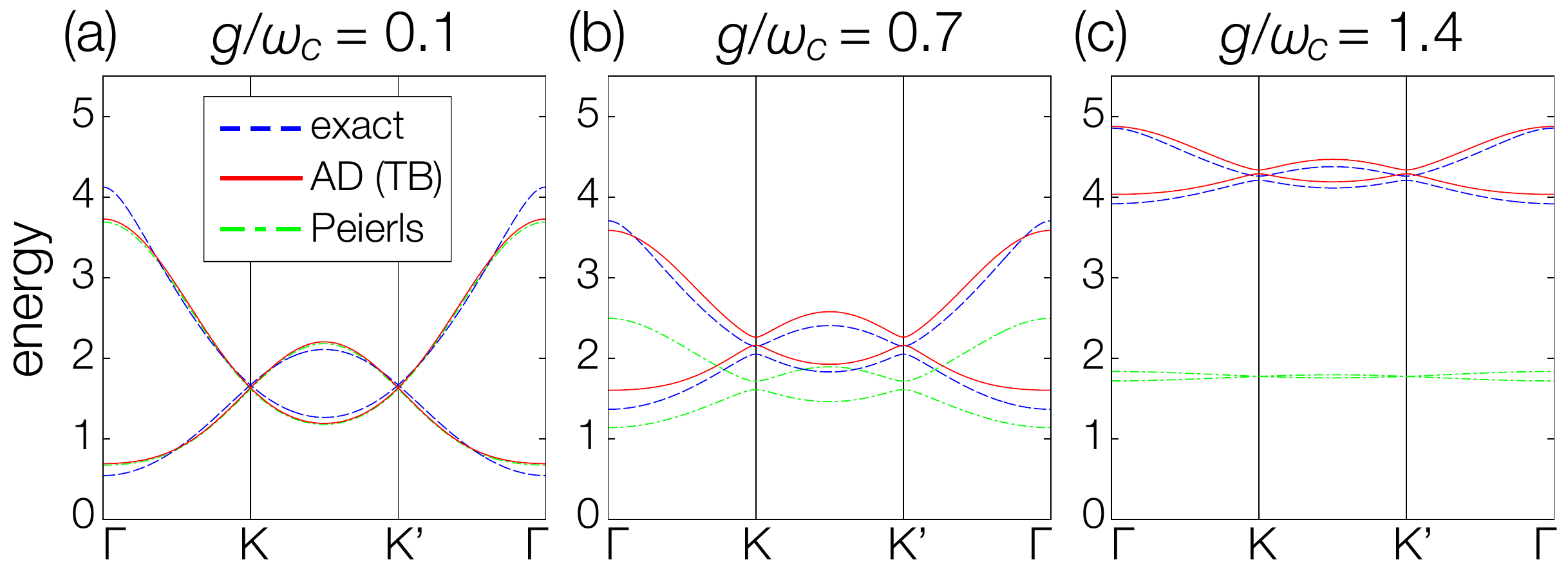}
  \caption{\label{figure4} Low-energy spectra at different coupling strengths obtained from the exact analysis of Eq.~\eqref{ham_AD} (blue), the tight-binding model~\eqref{ham_TB} in the AD frame (red), and the tight-binding model with Peierls substitution (green). K and K' denote the two Dirac points. Parameters are \(m\!=\!5, \omega_c\!=\!4, V_0\!=\!1.5, \Delta = 0\).}
\end{figure}

In Fig.~\ref{figure4}, we also show the low energy spectra at different coupling strengths. When the coupling \(g\) is increased, the Wannier orbitals in the AD frame become tightly localized around each potential minima due to the enhancement of the effective mass \(m_{\rm eff}\) in Eq.~\eqref{ham_mat}. Thus, effective hopping amplitudes $t_{ij}$ between different Wanneir orbitals are exponentially suppressed, leading to the flattening of energy bands (blue and red curves). Peierls substitution (green curves) tends to overestimate this flattening behavior especially in DSC regimes. We note that the upward shift of the entire energy spectrum is caused by the increase of the zero-point energy \(\hbar\Omega/2\) of cavity photons with the renormalized frequency \(\Omega\), which  cannot be captured by Peierls substitution. Similar behaviors can be also found at nonzero \(\Delta\).

We note that the minimally coupled Dirac Hamiltonian can be obtained by expanding the tight-binding Hamiltonian with Peierls substitutions around K and K' points. Thus, around these representative points, the low-energy spectrum evaluated from the minimally coupled Dirac Hamiltonian shows the similar behaviors as in the result of the Peierls substitution as far as we consider a weak coupling \(g/\omega_c\ll 1\). However, at large \(g\), these two results disagree with each other and, in particular, the minimally coupled Dirac Hamiltonian fails to reproduce the gap closing at K-point, which corresponds to the reentrant topological phase transition shown in Fig.~\ref{figure1}.

\section{Discussions and conclusions\label{sec:Discussion}}
We point out that our formalism reveals an intimate connection between quantum light-matter systems and the celebrated Haldane honeycomb model \cite{haldane_model_1988}, which allows us to develop a simple understanding of the present results. To see this, we first recall that low-energy physics of the present light-matter systems can be well captured by the tight-binding Hamiltonian, \(\sum_{\langle ij\rangle} t_{ij}|w_i\rangle\langle w_j|\), which only contains electron degrees of freedom. Here, as inferred from the TRS-breaking term \(\xi^2\bm p\!\times\!\bm e_z/(2\hbar)\) in  Eq.~\eqref{ham_mat}, the effective hopping amplitudes \(t_{ij}\) are in general complex-valued. In particular, the next-nearest neighbor hopping amplitudes \(t_2\) in the present model acquire phase factors, and the resulting tight-binding Hamiltonian becomes equivalent to the Haldane honeycomb model.
This analogy gives a simple interpretation of the reentrant TPT emerging in  DSC regimes (Fig.~\ref{figure1}). On the one hand, it is known in the Haldane model that reentrant transition to the trivial phase can occur when an imaginary part of \(t_2\) falls below the on-site energy difference $\sim\!\Delta$. On the other hand, in DSC regimes of the present light-matter model, the effective coupling $\xi$ (mass $m_{\rm eff}$) decreases (increases) as the coupling \(g\) is increased, resulting in the suppression of the imaginary part of \(t_2\). Since the on-site energy difference always remains to be \(\sim\!\Delta\), the reentrant transition is expected to occur in DSC regimes as found in Fig.~\ref{figure1}.

We expect that our analyses should be of relevance to recent experiments realizing strong light-matter interactions between low-dimensional materials and cavity photons \cite{keller_landau_2020,paravicini-bagliani_magneto-transport_2019,appugliese_breakdown_2022,orgiu_conductivity_2015,scalari_ultrastrong_2012,maissen_ultrastrong_2014,scalari_superconducting_2014,zhang_collective_2016}. Besides condensed matter systems, our consideration is also relevant to circuit QED architectures, which can simulate single-electron Hamiltonians with periodic potentials \cite{chirolli_enhanced_2021,rymarz_hardware-encoding_2021}. We expect that geometrical and topological phenomena revealed by the present analyses can be explored by coupling these experimental systems to a chiral cavity/resonator or waveguide \cite{owens_chiral_2022,hubener_engineering_2021}.

In summary, we developed a faithful and efficient theoretical framework to analyze geometry and topology in ultrastrongly coupled quantum light-matter systems. We applied our general approach to the concrete model of massive Dirac fermions on honeycomb lattice potential and revealed its complete phase diagram (Fig.~\ref{figure1}), featuring the cavity-induced topological phase transition and its reentrant behavior in genuinely nonperturbative regimes. The suppression of the topological phase in deep strong coupling regimes originates from intrinsic renormalizations of the electron mass and the effective coupling strength, and should be a general feature of cavity QED materials. We expect that our results are of relevance to a wide range of systems not only in cavity QED materials, but also in circuit QED \cite{blais_circuit_2021,gu_microwave_2017} and waveguide QED \cite{roy_colloquium_2017}.

\acknowledgments
We acknowledge support from the Japan Society for the Promotion of Science through Grant No.~JP19K23424.

\appendix

\section{Derivation of the transformed Hamiltonian for a single cavity mode with arbitrary polarization\label{App:AD_transformation}}
We here provide the derivation of the Hamiltonian~\eqref{ham_AD} obtained after the asymptotically decoupling (AD) unitary transformation in the case of a single cavity mode with arbitrary polarization. Specifically, we start from the general two-dimensional cavity QED Hamiltonian in the Coulomb gauge~\eqref{ham_Coulomb}:
\begin{align}
  \hat H_C    & = \frac{(\hat{\bm p}-q\hat{\bm A})^2}{2m} + V(\bm r) + \hbar\omega_c\left(\hat a^\dagger\hat a+\frac{1}{2}\right),\label{sm_ham_Coulomb} \\
  \hat{\bm A} & = A_0\left[(u\bm e+v\bm e^*)\hat a + (u^*\bm e^*+v^*\bm e)\hat a^\dagger\right],
\end{align}
where we expand the polarization vector \(\bm f\) in Eq.~\eqref{ham_Coulomb} by the circular polarization vectors \(\bm e = (1, -i)^{\rm T}/\sqrt{2}\) and \(\bm e^*=(1, i)^{\rm T}/\sqrt{2}\). Here, \(u\) and \(v\) are arbitrary complex numbers that satisfy \(|u|^2 + |v|^2 = 1\). We can take \(u\) and \(v\) to be real numbers without loss of generality; this is always possible by using a certain spatial rotation, \(p_{x(y)}\mapsto p_{x(y)} \cos\theta +\!(-) p_{y(x)} \sin\theta\), and a \(U(1)\) gauge transformation, \(\hat a\mapsto e^{i\alpha}\hat a\). We note that the case of the circular polarization discussed in the main text corresponds to the choice of \(u=1, v=0\).

To derive the expression of the transformed Hamiltonian, we first diagonalize the quadratic part, \(q^2\hat{\bm A}^2/2m + \hbar\omega_c(\hat a^\dagger\hat a + 1/2)\), via the Bogoliubov transformation and rewrite the Hamiltonian~\eqref{sm_ham_Coulomb} as
\begin{align}
  \hat H_C =    & \frac{\hat{\bm p}^2}{2m} + V(\bm r) + \hbar\Omega\left(\hat b^\dagger\hat b + \frac{1}{2}\right)\nonumber \\
                & - \frac{qA_0}{m}\hat{\bm p}\cdot
  \left(\begin{array}{c}
            \frac{u+v}{\sqrt{2}}\sqrt{\frac{\omega_{Ac}}{\Omega}}(\hat b^\dagger + \hat b) \\
            i\frac{u-v}{\sqrt{2}}\sqrt{\frac{\Omega}{\omega_{Ac}}}(\hat b^\dagger - \hat b)
          \end{array}\right),                            \\
  \omega_{Ac} = & (1-2uv)\omega_A + \omega_c,                                                                               \\
  \Omega =      & \sqrt{(\omega_A+\omega_c)^2 - (2uv\omega_A)^2},                                                           \\
  \omega_A =    & \frac{q^2A_0^2}{m\hbar},
\end{align}
where \(b^{(\dagger)}\) is a squeezed annihilation (creation) operator defined by
\begin{align}
  \hat b^\dagger + \hat b = & \sqrt{\frac{\Omega}{\omega_{Ac}}} (\hat a^\dagger + \hat a), \\
  \hat b^\dagger - \hat b = & \sqrt{\frac{\omega_{Ac}}{\Omega}} (\hat a^\dagger - \hat a).
\end{align}
We then use the unitary transformation \(\hat U\) defined by
\begin{align}
  \hat U         & = \exp\left( -ix_{\Omega} \frac{\hat{\bm p}}{\hbar}\cdot\hat{\bm \rho}\right),      \\
  x_{\Omega}     & = \sqrt{\frac{\hbar}{m\Omega}},                                                     \\
  \hat{\bm \rho} & =
  \left(\begin{array}{c}
            \hat \rho_x \\
            \hat \rho_y
          \end{array}\right)
  = \left(\begin{array}{c}
              \lambda_+ i \frac{\hat b^\dagger-\hat b}{\sqrt{2}} \\
              -\lambda_- \frac{\hat b^\dagger+\hat b}{\sqrt{2}}
            \end{array}\right),                            \\
  \lambda_{\pm}  & = \sqrt{\frac{\omega_A(1\pm 2uv)}{\omega_A(1\pm2uv)+ \omega_c}} {\rm sign}(u\pm v).
\end{align}
This unitary transformation \(\hat U\) acts on the electron position operator \(\bm r\) and the annihilation operator \(\hat b\) via
\begin{align}
  \hat U^\dagger \bm r\hat U   & = \bm r + x_{\Omega} \hat{\bm \rho} + \frac{\hbar\omega_A}{m\Omega^2}(u^2-v^2) \frac{\hat{\bm p}\times\bm e_z}{2\hbar}, \label{sm_ADtrans_r} \\
  \hat U^\dagger \hat b \hat U & = \hat b + \frac{x_{\Omega}}{\sqrt{2}\hbar}\left(\lambda_+ \hat p_x + i\lambda_- \hat p_y\right),
\end{align}
where the last term in the RHS of Eq.~\eqref{sm_ADtrans_r} originates from the noncommutativity between \(\hat \rho_x\) and \(\hat \rho_y\), i.e., \([\hat \rho_x,\hat \rho_y] = i(\omega_A/\Omega)(u^2-v^2)\). Using these relations, we finally arrive at the following transformed Hamiltonian \(\hat H^U = \hat U^\dagger\hat H_C\hat U\):
\begin{align}
  \hat H^U        = & \frac{\hat p_x^2}{2m_{x,{\rm eff}}} + \frac{\hat p_y^2}{2m_{y,{\rm eff}}} + \hbar\Omega\left(\hat b^\dagger\hat b + \frac{1}{2}\right)\nonumber                                            \\ & + V\left(\bm r + x_{\Omega} \hat{\bm \rho} + \frac{\hbar\omega_A}{m\Omega^2}(u^2-v^2) \frac{\hat{\bm p}\times\bm e_z}{2\hbar}\right),\label{ham_AD_gen} \\
  m_{x,{\rm eff}} = & \frac{\omega_A(1+2uv)+\omega_c}{\omega_c}, m_{y,{\rm eff}} =                                                                                    \frac{\omega_A(1-2uv)+\omega_c}{\omega_c},
\end{align}
which gives a generalization of Eq.~\eqref{ham_AD} in the main text. We note that the Hamiltonian in the AD frame~\eqref{ham_AD_gen} has a similar structure as Eq.~\eqref{ham_AD}. Thus, our formalism of evaluating physical observables and understandings of topological properties can also be applied to this case.

\section{Asymptotically decoupling unitary transformation for multiple circularly polarized cavity modes\label{App:AD_multi}}
We here extend the asymptotically decoupling (AD) unitary transformation used in the main text to the system with multiple circularly polarized cavity modes, which is described by the following Coulomb-gauge Hamiltonian:
\begin{align}
  \hat H_C    & = \frac{(\hat{\bm p}-q\hat{\bm A})^2}{2m} + V(\bm r) + \sum_{i=1}^{N}\hbar\omega_i\hat a^\dagger_i\hat a_i + \sum_{j=1}^M \hbar v_j\hat b^\dagger_j\hat b_j,\label{sm_ham_Coulomb_multi} \\
  \hat{\bm A} & = \sum_{i=1}^N f_i\left(\bm e\hat a_i + \bm e^*\hat a_i^\dagger\right) + \sum_{j=1}^M g_j\left(\bm e^*\hat b_j + \bm e\hat b_j^\dagger\right),
\end{align}
where \(\bm e = (1,-i)^{\rm T}/\sqrt{2}\), and \(\hat a_i^{(\dagger)}\) and \(\hat b_j^{(\dagger)}\) are the annihilation (creation) operators for the left and right circularly polarized modes with mode amplitudes \(f_i, g_j\in\mathbb{R}\) , respectively.

To diagonalize the quadratic part, \(\hat H_{\rm quad} = q^2\hat{\bm A}^2/2m + \sum_i\hbar\omega_i\hat a_i^\dagger\hat a_i + \sum_j\hbar v_j\hat b_j^\dagger\hat b_j\), we introduce the following conjugate pairs of variables,
\begin{align}
  \hat r_i = \left\{\begin{array}{ll}
                      \frac{\hat a_i^\dagger + \hat a_i}{\sqrt{2}}         & (i = 1,\ldots, N)             \\
                      \frac{\hat b_{i-N}^\dagger + \hat b_{i-N}}{\sqrt{2}} & (i = N\!+\!1,\ldots, N\!+\!M) \\
                    \end{array}\right., \\
  \hat q_i = \left\{\begin{array}{ll}
                      i\frac{\hat a_i^\dagger - \hat a_i}{\sqrt{2}}         & (i = 1,\ldots, N)             \\
                      i\frac{\hat b_{i-N}^\dagger - \hat b_{i-N}}{\sqrt{2}} & (i = N\!+\!1,\ldots, N\!+\!M) \\
                    \end{array}\right..
\end{align}
In this representation, the quadratic part can be expressed as
\begin{align}
  \hat H_{\rm quad} & = \frac{\hbar}{2} \hat{\bm r}^T A \hat{\bm r} + \frac{\hbar}{2} \hat{\bm q}^T B \hat{\bm q},   \\
  A                 & = \frac{q^2}{m\hbar} \bm w_1\bm w_1^T + {\rm diag}(\omega_1,\ldots,\omega_N, v_1,\ldots, v_M), \\
  B                 & = \frac{q^2}{m\hbar} \bm w_2\bm w_2^T + {\rm diag}(\omega_1,\ldots,\omega_N, v_1,\ldots, v_M),
\end{align}
where \(\hat{\bm r}\) and \(\hat{\bm q}\) denote \((\hat r_1,\ldots,\hat r_{N\!+\!M})^{\rm T}\) and \((\hat q_1,\ldots,\hat q_{N\!+\!M})^{\rm T}\), respectively. Here, we define the \(N\!+\!M\) dimensional vectors \(\bm w_1\) and \(\bm w_2\) by \(\bm w_1\!=\!(f_1,\ldots,f_N,g_1\ldots,g_M)\) and  \(\bm w_2\!=\!(f_1,\ldots,f_N,-g_1\ldots,-g_M)\). Since \(A\) and \(B\) are both symmetric and positive definite, there exists a \(N\!+\!M\) dimensional real matrix \(P\) that satisfies \(P^TAP = P^{-1}BP^{-T} = {\rm diag}(\Omega_1,\ldots,\Omega_{N\!+\!M})\). One can confirm this by decomposing \(P\) into \(P = B^{\frac{1}{2}}OD\), where \(O\) is an orthogonal matrix that diagonalizes \(B^{\frac{1}{2}}AB^{\frac{1}{2}}\), and \(D\) is a diagonal matrix defined by \(D = (O^TB^{\frac{1}{2}}AB^{\frac{1}{2}}O)^{-\frac{1}{4}}\). After replacing the conjugate pairs of variables via \(\hat{\bm r} = P\hat{\bm r}'\) and \(\hat{\bm q} = P^{-T}\hat{\bm q}'\), and introducing the corresponding annihilation (creation) operators \(\hat c_i^{(\dagger)} = (\hat r'_i+(-)i\hat q'_i)/\sqrt{2}\), we arrive at the following expression of the Coulomb gauge Hamiltonian~\eqref{sm_ham_Coulomb_multi}:
\begin{align}
  \hat H_C    & = \frac{\bm p^2}{2m}- \frac{q\hat{\bm A}}{m}\cdot\hat{\bm p} + V(\bm r) + \sum_{i=1}^{N\!+\!M} \hbar\Omega_i\hat c_i^\dagger\hat c_i, \\
  \hat{\bm A} & = \sum_{i=1}^{N\!+\!M} \left[(u_i\bm e + v_i\bm e^*)\hat c_i + (u_i\bm e^* + v_i\bm e)\hat c_i^\dagger\right],
\end{align}
where \(u_i\) and \(v_i\) are defined by \((u_1,\ldots,u_{N\!+\!M})^T  = \frac{1}{2}(P^T\bm w_1 + P^{-1}\bm w_2)\) and \((v_1,\ldots,v_{N\!+\!M})^T = \frac{1}{2}(P^T\bm w_1 - P^{-1}\bm w_2)\).  We introduce the AD transformation \(\hat U\)  by
\begin{align}
  \hat U           & = \exp\left( -i \frac{\hat{\bm p}}{\hbar}\cdot\sum_{i=1}^{N+M}\hat{\bm \rho}_i\right), \\
  \hat{\bm \rho}_i & = \frac{q}{m\Omega_i} (u_i\hat{\bm \pi}_i + v_i\hat{\bm \pi}'_i),
\end{align}
where \(\hat{\bm \pi}_i = -i\bm e\hat c_i + i\bm e^*\hat c_i^\dagger\) and \(\hat{\bm \pi}_i' =  -i\bm e^*\hat c_i + i\bm e\hat c_i^\dagger\). Using above relations, we finally obtain the following transformed Hamiltonian \(\hat H^U= \hat U^\dagger \hat H_C\hat U\):
\begin{align}
  \hat H^U    =   & \; \frac{\bm p^2}{2m_{\rm eff}} + \sum_{i=1}^{N+M}\hbar\Omega_i\hat c_i^\dagger\hat c_i\nonumber                                                                                                                           \\
                  & + V\left(\bm r + \sum_{i=1}^{N+M}\hat{\bm \rho}_i + \frac{\eta}{2\hbar}\bm p\times\bm e_z\right) ,    \label{ham_AD_gen_multi}                                                                                             \\
  m_{\rm eff} =   & m\left[ 1 + \frac{q^2}{m\hbar}\left(\sum_{i=1}^{N}\frac{f_i^2}{\omega_i} + \sum_{j=1}^{M}\frac{g_j^2}{v_j}\right)  \right],                                                                                                \\
  \eta          = & \frac{q^2}{m^2}\frac{\sum_{i=1}^N \frac{f_i^2}{\omega_i^2}-\sum_{j=1}^M \frac{g_j^2}{v_j^2}}{\left[ 1 + \frac{q^2}{m\hbar}\left(\sum_{i=1}^{N}\frac{f_i^2}{\omega_i} + \sum_{j=1}^{M}\frac{g_j^2}{v_j}\right)  \right]^2}.
\end{align}
Again, the Hamiltonian in the AD frame~\eqref{ham_AD_gen_multi} has a similar structure as Eq.~\eqref{ham_AD}, and our formalism and understandings of topological properties can be similarly applied.

\section{Derivation of the expressions for quantum geometrical quantities in the transformed frame\label{App:quantum_geometry}}
We next derive the expressions of quantum geometrical quantities, including  Berry curvature (see Eq.~\eqref{Berry_AD} in the main text), Berry connection, and quantum metric in the frame of reference after the AD unitary transformation. To this end, we first recall that the Bloch states in the Coulomb gauge can be related to the AD-frame counterparts by \(|\psi_{n\bm k}^C\rangle = \hat U|\psi_{n\bm k}^U\rangle\) with \(\hat U = \exp\left(-i\xi\frac{\bm p}{\hbar}\cdot\hat{\bm \pi}\right)\). Thus, the periodic parts of these Bloch states, \(|u_{n\bm k}^{C(U)}\rangle = e^{-i\bm k\bm r}|\psi_{n\bm k}^{C(U)}\rangle\), satisfy the following equality:
\begin{align}
  |u_{n\bm k}^C\rangle & = e^{-i\bm k\bm r}\hat Ue^{i\bm k\bm r}|u_{n\bm k}^U\rangle = \exp\left(-i\xi\frac{\bm p+\hbar\bm k}{\hbar}\cdot\hat{\bm \pi}\right) |u_{n\bm k}^U\rangle\label{sm_unitary_bewteen_uk}.
\end{align}
To take the derivatives of Eq.~\eqref{sm_unitary_bewteen_uk} with respect to \(k_x\) or \(k_y\), we must pay attention to the fact that the operators \(\hat \pi_x\) and \(\hat \pi_y\) do not commute with each other, but satisfy \([\hat \pi_x,\hat \pi_y] = i\). It is thus useful to first employ the Baker-Campbell-Hausdorff formula and bring the term containing \(k_x\) (or \(k_y\)) to the right of Eq.~\eqref{sm_unitary_bewteen_uk} as
\begin{widetext}
  \begin{align}
    |u_{n\bm k}^C\rangle & =  e^{-i\xi\frac{p_y+\hbar k_y}{\hbar}\hat\pi_y}e^{-i\xi\frac{p_x+\hbar k_x}{\hbar}\hat\pi_x} e^{-\frac{i\xi^2}{2\hbar^2}(p_x+\hbar k_x)(p_y+\hbar k_y)} |u_{n\bm k}^U\rangle \\
                         & = e^{-i\xi\frac{p_x+\hbar k_x}{\hbar}\hat\pi_x}e^{-i\xi\frac{p_y+\hbar k_y}{\hbar}\hat\pi_y} e^{\frac{i\xi^2}{2\hbar^2}(p_x+\hbar k_x)(p_y+\hbar k_y)} |u_{n\bm k}^U\rangle.
  \end{align}
  We can now take the derivatives of Eq.~\eqref{sm_unitary_bewteen_uk} with respect to \(k_x\) and \(k_y\) as follows:
  \begin{align}
    \partial_{k_x}|u_{n\bm k}^C\rangle & = \exp\left(-i\xi\frac{\bm p+\hbar\bm k}{\hbar}\cdot\hat{\bm \pi}\right)\left[-i\xi\hat\pi_x - \frac{i\xi^2}{2\hbar}(p_y+\hbar k_y) + \partial_{k_x}\right] |u_{n\bm k}^U\rangle,\label{sm_diff_kx} \\
    \partial_{k_y}|u_{n\bm k}^C\rangle & = \exp\left(-i\xi\frac{\bm p+\hbar\bm k}{\hbar}\cdot\hat{\bm \pi}\right)\left[-i\xi\hat\pi_y + \frac{i\xi^2}{2\hbar}(p_x+\hbar k_x) + \partial_{k_y}\right] |u_{n\bm k}^U\rangle\label{sm_diff_ky}.
  \end{align}
  The resulting expressions of Berry connection \(\bm A^C_n(\bm k) = i\langle u_{n\bm k}^C|\nabla_{\bm k} u_{n\bm k}^C\rangle\) and Berry curvature are
  \begin{align}
    \bm A^C_n(\bm k) & = i\langle u_{n\bm k}^U|\nabla_{\bm k}u_{n\bm k}^U\rangle + \xi\langle \psi_{n\bm k}^U|\hat{\bm \pi}|\psi_{n\bm k}\rangle + \frac{\xi^2}{2\hbar}\langle \psi_{n\bm k}^U|\hat{\bm p}\times\bm e_z|\psi_{n\bm k}\rangle,                       \label{sm_Berryconn}                                                                                                                                                  \\
    B^C_{n}(\bm k)   & = i\left(\langle\partial_{k_x} u_{n\bm k}^U|\partial_{k_y} u_{n\bm k}^U\rangle - \langle\partial_{k_y} u_{n\bm k}^U|\partial_{k_x} u_{n\bm k}^U\rangle\right)+ \xi \left[\bm \nabla_{\bm k}\!\times\!\langle\psi_{n\bm k}^U|\hat{\bm \pi}|\psi_{n\bm k}^U\rangle\right]_{k_z} \!\!-\! \frac{\xi^2}{2\hbar}\bm\nabla_{\bm k}\!\cdot\!\langle\psi_{n\bm k}^U|\hat{\bm p}|\psi_{n\bm k}^U\rangle.\label{sm_Berrycurv}
  \end{align}
  For the sake of completeness, we also provide the expression of the quantum geometric tensor \(\mathfrak{G}^C(\bm k)\) defined by
  \begin{align}
    \mathfrak{G}^C_{n,ij}(\bm k) & = \langle \partial_{k_i} u_{n\bm k}^C|\left[ 1-| u_{n\bm k}^C\rangle\langle u_{n\bm k}^C| \right]|\partial_{k_j} u_{n\bm k}^C\rangle,\ \ (i,j = x,y)\label{sm_quantum_geometric_tensor}.
  \end{align}
  We note that quantum metric \(g_{n,ij}^C(\bm k)\) is given by the real part of the quantum geometric tensor. Using Eqs.~\eqref{sm_diff_kx} and~\eqref{sm_diff_ky}, we can rewrite the quantum geometric tensor~\eqref{sm_quantum_geometric_tensor} in terms of the AD-frame Bloch states as
  \begin{align}
    \mathfrak{G}_{n,ij}(\bm k) = & \langle \partial_{k_i} u_{n\bm k}^U|\left[ 1-| u_{n\bm k}^U\rangle\langle u_{n\bm k}^U| \right]|\partial_{k_j} u_{n\bm k}^U\rangle                                                                                                                                               \nonumber                                      \\
                                 & + \langle u_{n\bm k}^U| \left(i\xi\hat\pi_i +\frac{i\xi^2}{2\hbar}\left[(\bm p+\hbar\bm k)\times\bm e_z\right]_i\right)\left[ 1-| u_{n\bm k}^U\rangle\langle u_{n\bm k}^U| \right]|\partial_{k_j} u_{n\bm k}^U\rangle                                                             \nonumber                                     \\
                                 & + \langle \partial_{k_i} u_{n\bm k}^U| \left[ 1-| u_{n\bm k}^U\rangle\langle u_{n\bm k}^U| \right] \left(-i\xi\hat\pi_j -\frac{i\xi^2}{2\hbar}\left[(\bm p+\hbar\bm k)\times\bm e_z\right]_j\right) | u_{n\bm k}^U\rangle                                                  \nonumber                                            \\
                                 & + \langle u_{n\bm k}^U |\left(i\xi\hat\pi_i +\frac{i\xi^2}{2\hbar}\left[(\bm p+\hbar\bm k)\times\bm e_z\right]_i\right)\left(-i\xi\hat\pi_j -\frac{i\xi^2}{2\hbar}\left[(\bm p+\hbar\bm k)\times\bm e_z\right]_j\right)| u_{n\bm k}^U\rangle                                      \nonumber                                     \\
                                 & - \langle u_{n\bm k}^U |\left(i\xi\hat\pi_i +\frac{i\xi^2}{2\hbar}\left[(\bm p+\hbar\bm k)\times\bm e_z\right]_i\right) | u_{n\bm k}^U\rangle \langle u_{n\bm k}^U | \left(-i\xi\hat\pi_j -\frac{i\xi^2}{2\hbar}\left[(\bm p+\hbar\bm k)\times\bm e_z\right]_j\right)| u_{n\bm k}^U\rangle.\label{sm_quantum_geometric_tensor2}
  \end{align}
\end{widetext}
In Eqs.~\eqref{sm_Berryconn}, \eqref{sm_Berrycurv}, and \eqref{sm_quantum_geometric_tensor2}, the first term in each equation gives the dominant contribution which can be efficiently evaluated by employing the electron-only tight-binding description~\eqref{Berry_AD_approx}. Nevertheless, we emphasize that it is also possible to include higher-order terms to systematically improve quantitative accuracy when necessary.

\section{Effects of electromagnetic fluctuations in the tight-binding model\label{App:EM_fluctuations}}
\begin{figure}[t]
  \centering
  \includegraphics[width = 8.5cm]{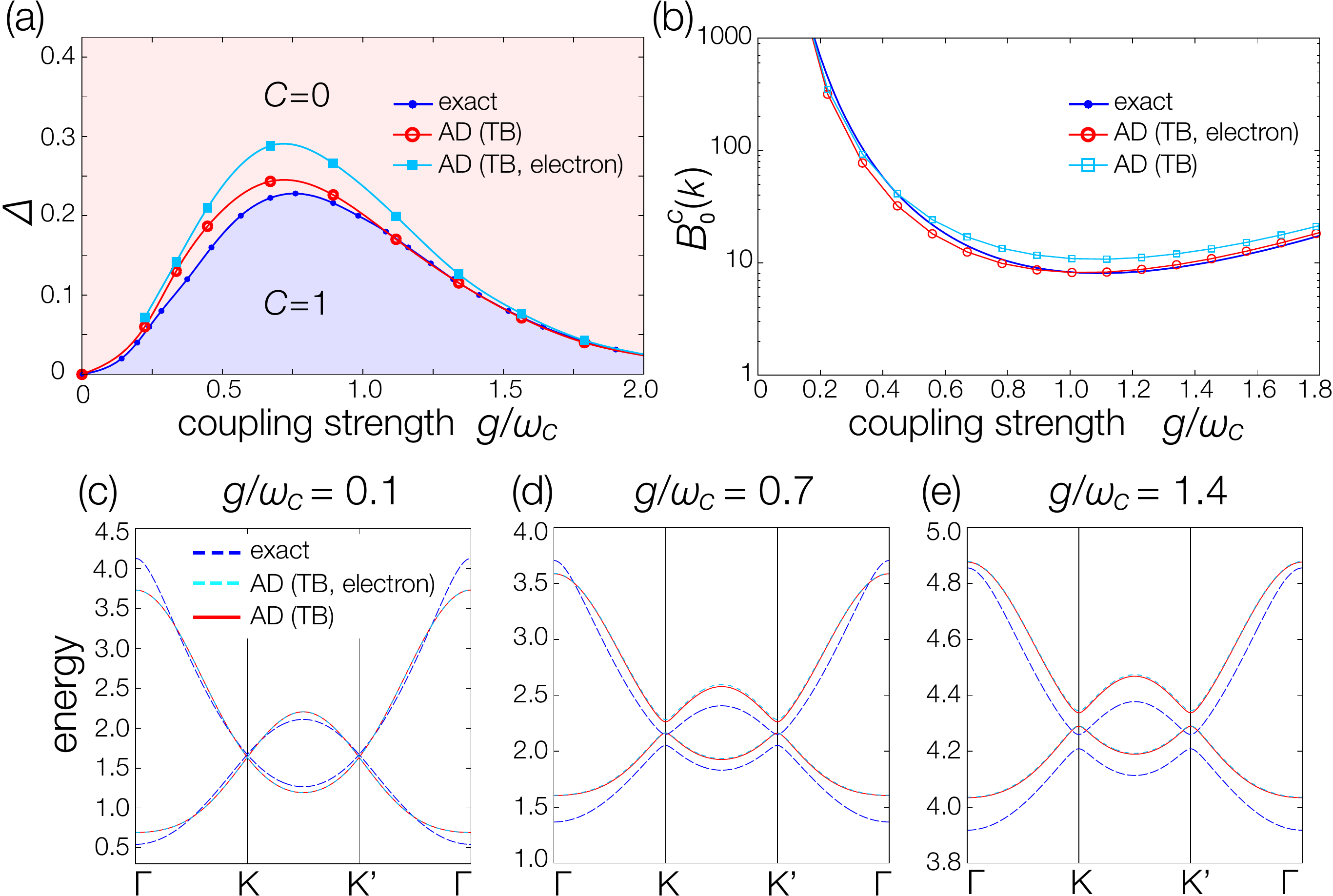}
  \caption{\label{sm_figure1}(a)-(e) Comparisons of the results obtained from the electron-photon tight-binding model in the transformed frame (Eq.~\eqref{ham_TB} in the main text) and its electron-part Hamiltonian~\eqref{sm_ham_TB_wo_EM}. Each panel corresponds to (a) phase boundary, (b) Berry curvature, and (c)-(e) energy spectra of the honeycomb model considered in the main text. Parameters are the same as in Figs.~\ref{figure1}, \ref{figure3}, and \ref{figure4} in the main text.}
\end{figure}

We here examine effects of electromagnetic fluctuations $\hat{\mu}_{ij}$ in the transformed frame by comparing the results of the electron-photon tight-binding Hamiltonian (Eq.~\eqref{ham_TB} in the main text) and its electron-only part,
\begin{align}
  \hat H_{{\rm TB},{\rm mat}}^U = \sum_{\langle ij\rangle} t_{ij}|w_i\rangle\langle w_j|.\label{sm_ham_TB_wo_EM}
\end{align}
Figure~\ref{sm_figure1} compares the results obtained from these two tight-binding Hamiltonians. Each panel corresponds to (a) phase boundary, (b) Berry curvature, and (c)-(e) energy spectra of the honeycomb model, respectively. The parameters are chosen in the same manner as in  Figs.~\ref{figure1}, \ref{figure3}, and \ref{figure4} in the main text. These comparisons demonstrate that the electromagnetic fluctuations in the transformed frame only lead to minor modifications, and qualitative features in the low-energy physics can be well captured simply by its electron-part Hamiltonian~\eqref{sm_ham_TB_wo_EM}, where the photon degrees of freedom are completely eliminated.

\section{Details of the conventional analysis using the minimally coupled Dirac Hamiltonian\label{App:minimally_coupled}}
We provide the technical details about the conventional analysis using the minimally coupled Dirac Hamiltonian considered in the main text. To derive the Dirac Hamiltonian, we start from the following honeycomb tight-binding model constructed from \(\hat H = {\bm p}^2/2m + V(\bm r)\) with the potential~\eqref{potential_massive_honeycomb} in the main text:
\begin{align}
  \hat H = & \sum_{i}\bigl(\varepsilon_A|w^0_{i,A}\rangle\langle w_{i,A}^0| + \varepsilon_B |w_{i,B}^0\rangle\langle w_{i,B}^0|\bigr)\nonumber                     \\
           & +  \sum_{\text{1st nearest neighbors}} t \bigl(|w_{i,A}^0\rangle\langle w_{j,B}^0| + |w_{j,B}^0\rangle\langle w_{i,A}^0|\bigr),\label{sm_ham_tb_bare}
\end{align}
where \(|w^0_{i,A(B)}\rangle\) denotes the Wanneir orbital localized at the sublattice A (B) in the \(i\) th unit cell. Here, \(\varepsilon_{A(B)}\) is the on-site energy of the sublattice A(B), and \(t\) is the hopping amplitude to the first nearest neighbors (see also Fig.~\ref{sm_figure2} below). In the basis of two-component spinors \((|\psi_{\bm k,A}\rangle, |\psi_{\bm k,B}\rangle)\) corresponding to the Bloch states on the two sublattices, this tight-binding Hamiltonian becomes
\begin{align}
  \hat H(\bm k) = & \frac{\varepsilon_A + \varepsilon_B}{2} \hat I_2 + \frac{\varepsilon_A-\varepsilon_B}{2} \hat\sigma_z\nonumber                      \\
                  & + \sum_{i = 1,2,3} \left[ \cos(\bm k\cdot\bm \tau_i)\hat\sigma_x - \sin(\bm k\cdot\bm \tau_i)\hat\sigma_y \right],\label{sm_ham_Hk}
\end{align}
where \(\hat\sigma_i\) (\(i = x,y,z\)) are the Pauli matrices and \(\bm \tau_i\) (\(i=1,2,3\)) are the three displacement vectors from A site to its three nearest-neighbor B sites. By expanding Eq.~\eqref{sm_ham_Hk} around the two Dirac points (\(\bm K = \pm(\bm b_1/3 + \bm b_2/3)\)), we obtain the following massive Dirac Hamiltonians:
\begin{align}
  \hat H_{\rm Dirac}(\bm k) = & \; \frac{\varepsilon_A + \varepsilon_B}{2} \hat I_2 + \frac{\varepsilon_A-\varepsilon_B}{2} \hat\sigma_z \nonumber \\
                              & + \frac{\sqrt{3}t}{2}(k_x\hat\sigma_x \pm k_y\hat\sigma_y).\label{sm_ham_Dirac}
\end{align}
The minimally coupled Dirac Hamiltonian \(\hat H^A_{\rm Dirac}(\bm k)\) can then be obtained after replacing the wavevector \(\bm k\) in Eq.~\eqref{sm_ham_Dirac} by \(\bm k - (q/\hbar)\hat{\bm A}\):
\begin{align}
  \hat H^A_{\rm Dirac}(\bm k) = & \; \frac{\varepsilon_A + \varepsilon_B}{2} \hat I_2 + \frac{\varepsilon_A-\varepsilon_B}{2} \hat\sigma_z+ \hbar\omega_c\hat a^\dagger\hat a \nonumber                   \\
  +                             & \frac{\sqrt{3}t}{2}\left(\left(k_x-\frac{q\hat A_x}{\hbar}\right)\hat\sigma_x \pm \left(k_y-\frac{q\hat A_y}{\hbar}\right)\hat\sigma_y\right) ,\label{sm_ham_min_Dirac}
\end{align}
where \(\hat{\bm A}\) is the vector potential operator defined by Eq.~\eqref{vector_potential} in the main text. In the phase diagram of the  honeycomb model (Fig.~\ref{figure1} in the main text), we determine the phase boundary of the minimally coupled Dirac Hamiltonian~\eqref{sm_ham_min_Dirac} by locating the coupling \(g\) at which the energy gap above the lowest electron-polariton band closes. We also evaluate the Chern number (inset of Fig.~\ref{figure1} in the main text) by integrating Berry curvature of the lowest electron-polariton band over a sufficiently large area and summing up the contributions from the two Dirac points.

\begin{figure}[t]
  \centering
  \includegraphics[width = 8cm]{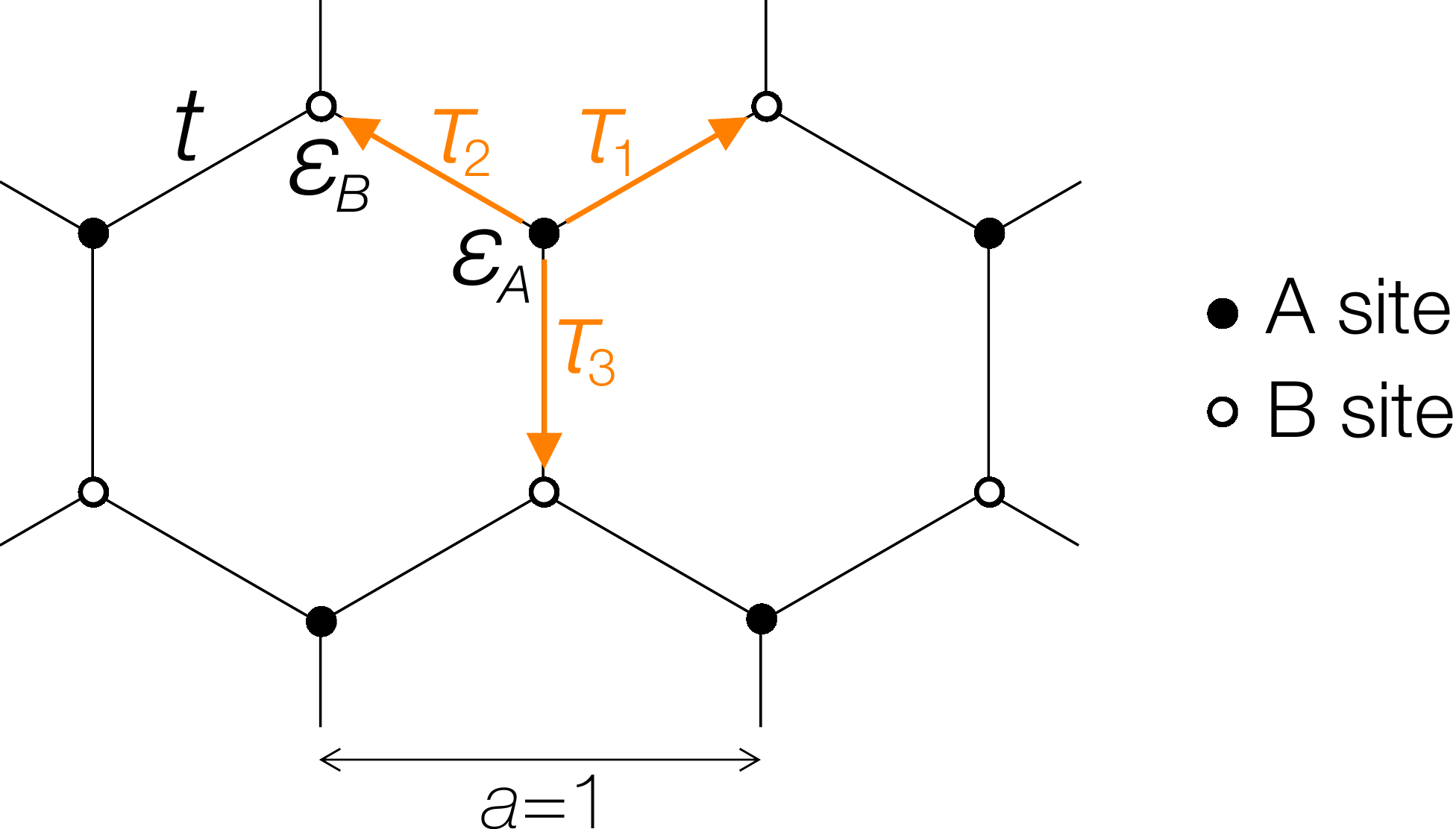}
  \caption{\label{sm_figure2}Honeycomb tight-binding model in Eq.~\eqref{sm_ham_tb_bare}. Solid and open points represent A- and B-sublattice sites, respectively. Here, \(\varepsilon_{A(B)}\) is the on-site energy of each sublattice A(B), and \(t\) is the hopping amplitude to the nearest neighbors.}
\end{figure}

\section{Details of the exact analysis\label{App:exact}}
We here provide the technical details about the exact analysis performed in the main text. To obtain exact results at ultrastrong coupling strengths, we use the continuum Hamiltonian in the AD frame~\eqref{ham_AD} as
\begin{align}
  \hat H^U = \; & \frac{\hat{\bm p}^2}{2m_{\rm eff}} + \hbar\Omega\left(\hat a^\dagger\hat a + \frac{1}{2}\right)\nonumber                             \\
                & + \sum_{\bm G} V_{\bm G} e^{i\bm G\bm r}e^{\frac{i\xi^2}{2\hbar}\bm G\cdot\hat{\bm p}\times\bm e_z} e^{i\xi\bm G\cdot\hat{\bm \pi}}.
\end{align}
The matrix elements of \(\hat H^U_{\bm k} \equiv e^{-i\bm k\bm r}\hat H^Ue^{i\bm k\bm r}\) with respect to \(|\bm K\!\otimes\!n\rangle\), which is the tensor product of the plane wave with wavevector \(\bm K\in \mathbb{Z}\bm b_1 + \mathbb{Z}\bm b_2\) and photon number eigenstate \(|n\rangle\), are given by
\begin{align}
   & \langle \bm K\otimes n|\hat H^U_{\bm k}|\bm K'\otimes n'\rangle \nonumber                                                                                                   \\ &= \frac{\hbar^2(\bm K+\bm k)^2}{2m_{\rm eff}}\delta_{\bm K,\bm K'} + \hbar\Omega\left(n + \frac{1}{2}\right)\delta_{nn'} \nonumber           \\
   & \;\;\;\;\; + \sum_{\bm G} V_{\bm G} e^{\frac{i\xi^2}{2}(\bm G\times(\bm K+\bm k))_z} \delta_{\bm K,\bm K'+\bm G}\times \langle n|e^{i\xi\bm G\cdot\hat{\bm \pi}}|n'\rangle.
\end{align}
We can perform the exact diagonalization of \(\hat H_{\bm k}^U\) to obtain its spectrum and eigenstate \(|u_{n\bm k}^U\rangle\) at the expense of high computational costs.

To evaluate the Berry curvature~\eqref{Berry_AD}, we start with the equivalent expression~\eqref{Berry_Coulomb} in the Coulomb gauge. We first rewrite it as
\begin{align}
  B_{n}^C(\bm k) & \!=\! -2{\rm Im} \sum_{m\neq n} \frac{\langle u_{n\bm k}^C|\partial_{k_x}\hat H_{\bm k}^C|u_{m\bm k}^C\rangle\langle u_{m\bm k}^C|\partial_{k_y}\hat H_{\bm k}^C|u_{n\bm k}^C\rangle}{(\varepsilon_{n\bm k}-\varepsilon_{m\bm k})^2},\label{sm_Berry_Coulomb}
\end{align}
where \(\varepsilon_{n\bm k}\) is the \(n\)-th eigenvalue of \(\hat H_{\bm k}^C\!\equiv\!e^{-i\bm k\bm r}\hat H_{C}e^{i\bm k\bm r}\). Since \(|u_{n\bm k}^C\rangle\) and \(|u_{n\bm k}^U\rangle\) are related via Eq.~\eqref{sm_unitary_bewteen_uk}, we can rewrite \(\langle u_{n\bm k}^C|\nabla\hat H_{\bm k}^C|u_{m\bm k}^C\rangle\) in Eq.~\eqref{sm_Berry_Coulomb} in terms of Bloch states in the AD frame as
\begin{align}
  \langle u_{n\bm k}^C|\nabla \hat H_{\bm k}^C|u_{m\bm k}^C\rangle
   & = \hbar\langle u_{n\bm k}^C|\left(\frac{\hat{\bm p}+\hbar\bm k-q\hat{\bm A}}{m}\right)|u_{m\bm k}^C\rangle                                              \\
   & = \hbar\langle u_{n\bm k}^U|\left(\frac{\hat{\bm p}+\hbar\bm k}{m_{\rm eff}}-\frac{q\hat{\bm A}}{m}\right)|u_{m\bm k}^U\rangle.\label{sm_Berry_matelem}
\end{align}
With Eqs.~\eqref{sm_Berry_Coulomb} and \eqref{sm_Berry_matelem}, we obtain the following expression of Berry curvature \(B_{n}^C(\bm k)\), which is used in evaluating the exact Berry curvature in Fig.~\ref{figure3}:
\begin{widetext}
  \begin{align}
    B_n^C(\bm k) & = -2\hbar^2 {\rm Im} \sum_{m\neq n} \frac{\langle u_{n\bm k}^U|\left( \frac{\hat{p}_x+\hbar k_x}{m_{\rm eff}}-\frac{q\hat{A}_x}{m} \right)|u_{m\bm k}^U\rangle\langle u_{m\bm k}^U|\left( \frac{\hat{p}_y+\hbar k_y}{m_{\rm eff}}-\frac{q\hat{A}_y}{m} \right)|u_{n\bm k}^U\rangle}{(\varepsilon_{n\bm k}-\varepsilon_{m\bm k})^2}\label{sm_Berry_exact}.
  \end{align}
\end{widetext}
\bibliography{Cavity_Bloch_theory}
\end{document}